\documentclass{aa}  
\usepackage{graphicx}
\usepackage{txfonts}

\def\teff{${\rm T_{\rm eff}}$}
\def\gr{$\log g$}
\def\fei{[Fe~I/H]}
\def\feii{[Fe~II/H]}
\def\ali{${\rm A(\rm Li})_{\rm NLTE}$}
\def\kms{${\rm km~s^{-1}}$}
\def\ciso{${\rm ^{12}C/^{13}C}$}

\begin{document}

\title{{\sl Curiouser and curiouser}:  the peculiar chemical composition of the Li/Na-rich star in
$\omega$ Centauri
\thanks{Based on observations collected at the ESO-VLT under programs 060.A-9700, 096.D-0728, 099.D-0258 and 0101.D-0620.}}

\author{A. Mucciarelli\inst{1,2}, L. Monaco\inst{3}, P. Bonifacio\inst{4}, M. Salaris\inst{5}, 
I.Saviane\inst{6}, B. Lanzoni\inst{1,2}, Y. Momany\inst{7}, G. Lo Curto\inst{6}}
\offprints{A. Mucciarelli}
\institute{
Dipartimento di Fisica e Astronomia {\sl Augusto Righi}, Universit\`a degli Studi di Bologna, Via Gobetti 93/2, I-40129 Bologna, Italy;
\and
INAF - Osservatorio di Astrofisica e Scienza dello Spazio di Bologna, Via Gobetti 93/3, I-40129 Bologna, Italy;
\and
Departamento de Ciencias Fisicas, Universidad Andres Bello, Fernandez Concha 700, Las Condes, Santiago, Chile
\and
GEPI, Observatoire de Paris, PSL Research University, CNRS, Place Jule Janssen 92190, Meudon, France
\and
Astrophysics Research Institute, Liverpool John Moores University, 146 Brownlow Hill, Liverpool L3 5RF, United Kingdom 
\and
European Southern Observatory, Alonso de Cordova 3107, Vitacura, Santiago, Chile
\and
INAF - Osservatorio Astronomico di Padova, Vic. dell'Osservatorio 5, 35122 Padova, Italy
}

\authorrunning{A. Mucciarelli et al.}
\titlerunning{Li-Na-rich star in $\omega$ Centauri}

\date{Submitted to A\&A }
 
\abstract
{We present a multi-instrument spectroscopic analysis of the unique Li/Na-rich giant star 
\#25664 in $\omega$ Centauri using spectra acquired with FLAMES-GIRAFFE, X-SHOOTER, UVES and HARPS. 
Li and Na abundances have been derived from the UVES spectrum using transitions 
weakly sensitive to non-local thermodynamic equilibrium and assumed isotopic ratio. 
This new analysis confirms the surprising Li and Na abundances of this star
(\ali\ =+2.71$\pm$0.07 dex, ${\rm [Na/Fe]}_{\rm NLTE}$=+1.00$\pm$0.05 dex).
Additionally, we provide new pieces of evidence for its chemical characterisation.
The \ciso\ isotopic ratio  (15$\pm$2) shows that this star 
has not yet undergone the extra-mixing episode usually 
associated with the red giant branch bump. Therefore, we can rule out the scenario of 
efficient deep extra-mixing during the red giant branch phase envisaged 
to explain the high Li and Na abundances.
Also, the star exhibits high abundances of both C and N ([C/Fe]=+0.45$\pm$0.16 dex 
and [N/Fe]=+0.99$\pm$0.20 dex),  not compatible with the typical C-N anticorrelation observed in globular cluster stars.
We found evidence of a radial velocity variability in \#25664, suggesting that the star could be part of a binary system, 
likely having accreted material from a  more massive companion when the latter was evolving in the AGB phase. 
Viable candidates for the donor star are AGB stars with 3-4 ${\rm M_{\odot}}$ 
and super-AGB stars ($\sim$7-8 ${\rm M_{\odot}}$), both able to produce Li- and Na-rich material. 
Alternatively, the star could be formed from the pure ejecta of a super-AGB stars, 
before the dilution with primordial gas occurs.}


   \maketitle
%

\section{Introduction}

The observed run of A(Li)\footnote{A(Li)=$\log{\frac{N_{Li}}{N_{H}}}+12$} 
in field halo and globular clusters between [Fe/H]$\sim$--3.0 and --1.0 dex
as a function of the stellar luminosity (or the surface gravity), 
is characterised by two {\sl plateaus} and two drops
\citep[see e.g.][]{gratton00,lind09,msb12}. 
These drops are driven by the mixing episodes occurring during the evolution of 
low-mass stars.
Upper main sequence stars exhibit a constant value of A(Li) ($\sim$2.1-2.3 dex), 
regardless of the effective temperature (\teff ) and metallicity,   
the so-called {\sl Spite Plateau} \citep{spite82,rebolo88,bm97}, originally interpreted 
as the signature of the Li abundance produced during the Big Bang nucleosynthesis. 

When the surface convection reaches regions hotter than $\sim2.5\cdot10^6$ K 
where Li burns, Li-free material is dredged to the surface (first dredge-up) 
with the consequent reduction of the photospheric A(Li) by $\sim$1.2~dex. 
After convection has attained its maximum depth at the end of the first dredge-up,  
stars fainter than the red giant branch (RGB) bump show a constant A(Li)$\sim$1.0 dex \citep{msb12}. 
The extra-mixing episode associated to the RGB bump leads to a subsequent depletion of 
the surface A(Li) that totally disappears from the stellar atmosphere \citep{charb07}.

Li-rich stars are peculiar stars that contradict this framework, 
exhibiting surface A(Li) 
significantly higher (up to 3 dex) than those measured in stars of similar luminosity. 
These stars are rare and they have been detected among all the evolutionary sequences, 
with evidence of a higher incidence during the phase of red clump 
\citep{casey19,kumar20}.
The origin of Li-rich stars is still debated and
the most popular scenarios envisaged to explain the enhancement of A(Li) are: \\
{\sl (i)}~engulfment of planets or brown dwarfs \citep{siess,aguileira16,casey16,aguilera20} that should increases 
the surface abundances of Be, $^{6}$Li and $^{7}$Li (even if the original chemical abundances could be restored 
after a given time). 
This scenario is favoured in metal-rich stars with [Fe/H]$>$--0.5 dex 
\citep[see][]{j10pl,casey16}
and in stars brighter than the RGB bump, because the larger stellar radius favours the engulfment process;\\
{\sl (ii)}~internal production of fresh Li through the Cameron-Fowler mechanism \citep{cf71}. This mechanism can occur during the RGB phase if deep extra mixing is able to circulate  matter  between  the  base  of  the  convective  envelope  and  a  region  close  to  the  H-burning  shell \citep{booth95,booth99,dv03}.
Also, the Cameron-Fowler mechanism can occur during the phase of asymptotic giant branch (AGB), in particular a significant production of Li is expected for the so-called super-AGB stars, with masses between 7 and 8 ${\rm M_{\odot}}$ , during 
the hot bottom burning phase \citep{ventura11,doherty14};\\
{\sl (iii)}~external production of fresh Li: the measured over-abundance of Li is the result of a mass transfer process in a binary system, from a companion that had produced  Li through the mechanisms described above.

The interpretation of Li-rich stars in globular clusters (GCs) 
is complicated by the observed abundance (anti-)correlations
among the light elements (i.e. He, C, N, O, Na, Mg, Al) 
involved in the hot CNO-cycles, usually interpreted as the outcome of a self-enrichment process 
occurring in the early stage of life of the cluster \citep{bastian18,gratton19}. 
As a crude classification, we are used to divide the cluster stars in two groups according to their light element abundances: 
first population (1P) stars  with abundance ratios that resemble those measured in the field stars 
with metallicity similar to that of the cluster, and 
second population (2P) stars showing in their chemical composition the signature of the hot CNO-cycle 
(in particular, the Na-O anticorrelation observed in all the old GCs).

The hot CNO-cycle producing the chemical anomalies observed in GCs occurs at temperatures higher than $10^7$K, one order of magnitude higher than that of Li-burning.
Therefore 2P stars should be Li-free or at least show a significant difference in A(Li) with respect to 1P stars.
Surprisingly, GCs with Li measurements display only small differences between A(Li) in 1P and 2P stars \citep{pasquini05,ghb09,lind09,monaco12,dobrov14} with 
the only remarkable exception of the multi-iron GC-like system $\omega$ Centauri \citep{johnson10,marino11} 
that exhibits an extend Li-Na anticorrelation \citep{mu18}.

Until now, only 15 Li-rich stars have been discovered in GCs 
\citep[see Fig.~3 in][and references therein]{sanna20}. 
Among them, one of the two Li-rich stars discovered in $\omega$ Centauri by \citet{mu19} is particularly intriguing.
The star \#25664 is a lower RGB (LRGB), fainter than the RGB bump luminosity level. Its membership has been 
confirmed according to its measured 
radial velocity \citep{mu19} and proper motions \citep{bellini09,brown21}

It shows an enhancement of Li (\ali\ =+2.40$\pm$0.06 dex) coupled 
with an extraordinarily high abundance of Na. The abundance from the Na I D lines is [Na/Fe]=+0.87$\pm$0.07 dex \citep{mu19} which is about 0.5 dex higher than the Na content measured in the most Na-rich stars 
in $\omega$ Centauri by \citet{mu18} using the same Na lines. 
These two uncommon chemical abundances make \#25664 a peculiar and unique object, without similar counterparts 
in other GCs or among the field stars.
As discussed by \citet{mu19}, the enhancement of Li and Na in this star is compatible with both an internal production during the RGB phase (if the star experienced efficient deep mixing) and with Li production in super-AGB stars. In the latter case, \#25664 could be formed directly from the pure ejecta of a super-AGB star \citep{dantona12} 
or it could be member of a former binary system experiencing mass transfer of Li/Na-rich material 
from the massive companion, when the latter was in the super-AGB phase. 

In this paper we present new spectroscopic observations of this star,
measuring the abundances of Na, Li but also C, N, O, Mg, Al and K,
to provide a complete view of its chemical composition 
and try to explain its origin. The paper is organised as follows: Section 2 describes the observations, 
Section 3 describes the measure of the radial velocities (RV), Section 4 presents the chemical analysis and 
the derived abundances, Section 5 discusses the results.

\section{Observations}
A multi-instrument spectroscopic campaign has been performed to properly characterise the kinematics 
and chemistry of the peculiar Li/Na-rich star \#25664 in $\omega$ Centauri. The acquired spectroscopic 
datasets are the following (see Table~\ref{rvtab}):
\begin{itemize}
\item {\sl FLAMES-GIRAFFE} \citep{pasquini02}: 
we secured one exposure with the setup HR12, two with HR13 and three with HR15N observed between 2016 February 1 and 19 
(Program ID: 096.D-0728, PI: Mucciarelli). These observations are described and discussed in \citet{mu18,mu19}.
\item {\sl X-SHOOTER} \citep{vernet11}: X-SHOOTER spectra of \#25664 and of two comparison LRGB stars 
in $\omega$ Centauri have been secured during the nights 2017 June 16 and 17 (Program ID:099.D-0258, PI: Origlia), 
adopting slits of 1.0'' and 0.9'' for the UVB and VIS arms, respectively, corresponding to spectral resolutions 
of 5,400 and 8,900.
The two comparison LRGB stars have been selected from the sample by \cite{mu18} with  atmospheric parameters 
and metallicity very similar to those of \#25664 but different Na abundances, hence belonging to 
different cluster populations. The two stars are \#77093 (1P) with [Na/Fe]=--0.29 dex and
\#329049 (2P) with [Na/Fe]=~+0.12 dex \citep{mu18}. 
The spectrum of \#25664 has been obtained with an exposure time of 1800 sec, while 1200 sec of integration 
have been used for both the comparison stars. The observations have been performed in nodding mode. 
Signal-to-noise (S/N) ratio per pixel is about 80-100 for the UBV arm and 50-60 for the VIS arm.
\item {\sl UVES} \citep{dekker00}: two exposures of 1800 sec and one exposure of 1200 sec have been 
obtained with UVES in the Dichroic mode and a slit of 1'' (providing a spectral resolution of about 40,000), 
adopting the Blue Arm setting CD\#1 390 (ranging from 3280 to 4490 \AA ) and the Red Arm setting CD\#3 580 
(ranging from 4800 to 6800 \AA ), during the nights 2018 June 7 and 8 (Program ID:0101.D-0620, PI: Ferraro). 
The S/N ratio per pixel around the Li line is about 65.
\item {\sl HARPS} \citep{mayor03}: the HARPS spectrograph provides a spectral resolution of 115,000 
(in HAM mode) and 80,000 (in EGSS mode), ranging from 3780 to 6910 \AA\ .
In HAM mode, three exposures of 1200 sec each have been acquired during the night 2018 June 26, three exposures of 1200 sec each 
during the night 2018 July 24, and one exposure of 3600 sec during the night 2019 February 20. 
One exposure of 3600 sec in EGGS mode has been secured during the night 2021 January 26.
 Multiple exposures of the same night have been co-added together.
The S/N ratio per pixel around the Li line of these spectra ranges from about 5 to 10-11.
\end{itemize}

\section{Radial velocities}

For \#25664 a RV of +227.8$\pm$0.2 \kms\ has been provided by \citet{mu19} 
as average of RVs derived from the six individual GIRAFFE spectra.
RVs from X-SHOOTER, UVES and HARPS spectra have been derived by cross-correlating them against appropriate 
synthetic template spectra calculated with the {\tt SYNTHE} code \citep{sbordone04,kurucz05} and convoluted 
with a Gaussian profile to reproduce the instrumental broadening. The cross-correlation 
has been performed with the {\tt IRAF}\footnote{IRAF is distributed by the National Optical Astronomy Observatory, which is operated by the Association of Universities for Research inAstronomy (AURA) under a cooperative agreement with the NationalScience Foundation.} task {\tt FXCOR}. 
We checked the accuracy of the zero-point of the wavelength calibration for each spectrum by 
cross-correlating emission/absorption telluric features against a synthetic spectrum 
of the Earth atmosphere calculated with {\tt TAPAS} \citep{bertaux14}. 
Only for the X-SHOOTER spectrum we found a small offset of +1.3 \kms\ that we accounted for.

Uncertainties in the RV measurement (from both photospheric and telluric lines) due to the photon noise, 
the spectral resolution and the finite size of the pixel have been estimated by using 
Monte Carlo simulations. We added Poisson noise to a synthetic spectrum calculated 
with the stellar parameters of \#25664 and to a Earth atmosphere synthetic spectrum, with the spectral 
resolution and the pixel size corresponding to each instrument. 
For each individual spectrum, a sample of 200 synthetic noisy spectra has been generated and
analysed as done for the observed spectra. 
The dispersion of the derived RV distribution has been taken as 
1$\sigma$ uncertainty in the RV measure. We added in quadrature the uncertainty 
related the measure of RVs in emission/absorption telluric features.

\begin{figure} 
\centering
\includegraphics[clip=true,scale=0.45,angle=0]{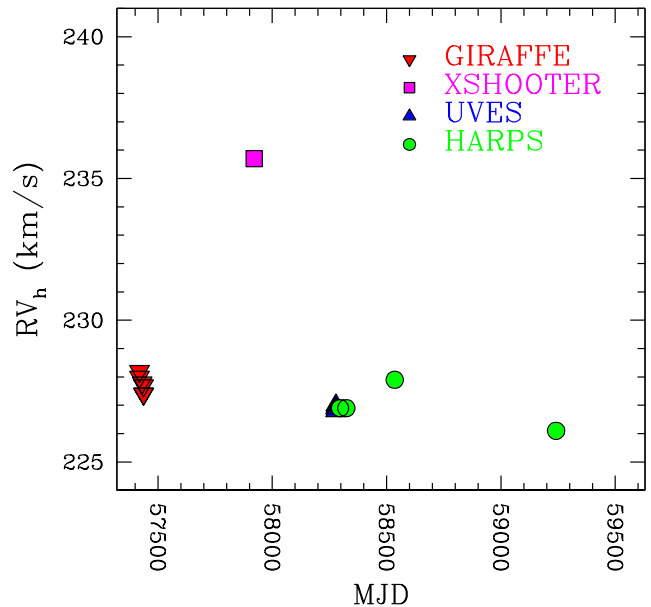}
\caption{Behaviour of heliocentric RV as a function of the modified Julian date derived 
from multi-instrument spectroscopic dataset listed in Tab.~\ref{rvtab}. 
Errors on RV are smaller than the symbols.}
\label{rvall}
\end{figure}

All the individual heliocentric RVs are listed 
in Table~\ref{rvtab}, while Fig.~\ref{rvall} shows their behaviour 
as a function of the modified Julian date. All the measures are distributed over a range of about 2 \kms\  
but that from X-SHOOTER spectrum that is higher by $\sim$7-8 \kms\ than the other RVs. 
Even if the RVs measured from X-SHOOTER spectra are the most uncertain,
we have no reason to exclude or consider less reliable the X-SHOOTER RV. 
To support the validity of the RV derived from X-SHOOTER, we checked the RVs obtained for the two comparison stars 
with those measured by \citet{mu18} from the GIRAFFE spectra, finding a difference (${\rm RV_{XSH}}$-${\rm RV_{GIR}}$) 
of --0.8 \kms\ for \#77093 and --0.5\kms\ for \#329049. 
The excellent agreement between the RVs from  X-SHOOTER and GIRAFFE spectra 
for these two stars suggests that the RV of \#25664 from the X-SHOOTER spectrum 
is correct or at least not affected from relevant biases.

\begin{figure} 
\centering
\includegraphics[clip=true,scale=0.45,angle=0]{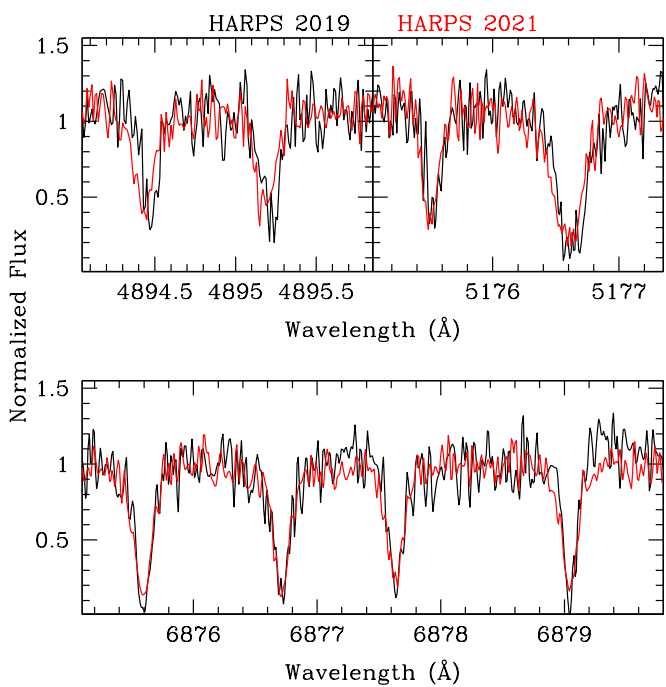}
\caption{Upper panels: portions of the HARPS spectra around some photospheric lines 
acquired in 2019 February (black line) 
and 2021 January (red line). The two spectra are corrected for the corresponding 
heliocentric correction. Lower panel: portions of the same HARPS spectra around telluric lines,
without applying the heliocentric correction.}
\label{harps}
\end{figure}

Also, we found a difference of 1.8 \kms\ between the RVs from the last two 
HARPS spectra. This difference is confirmed also when we consider RVs derived 
with the HARPS pipeline (providing an offset of 1.7 \kms\ between the two epochs).
Among the used instruments, HARPS provides the most reliable 
RVs because its high spectral resolution and the excellent stability of the 
instrument. The real difference in the RVs of these two epochs can be appreciated 
in Fig.~\ref{harps}. The upper panels show portions of the two spectra (both 
corrected for the corresponding heliocentric correction)
where the displacement between the lines is clearly visible. 
On the other hand, the position of telluric lines of the two spectra 
(without applying the heliocentric correction)
shown in the lower panel demonstrates that these spectra are 
perfectly aligned each other. 
We conclude that the star \#25664 exhibits clear evidence of RV variability suggesting that it is 
member of a binary system.
We also investigated the eventual photometric variability of   \#25664 within  the long-term CCD-monitoring of 
\citet{momany2020}. In this latter, the authors collected a wide-field ($\omega$CAM@VST) almost-simultaneous $u/r$-sloan monitoring of selected GCs that, in the case of $\omega$ Centauri, covered almost $\sim4$\,years. The $u/r$-light curves of \#25664  basically confirm that the star does not show any significant photometric variability (at least within a $\sim3\sigma$-level of their achieved photometric precision).

\begin{table*}
\caption{Observing log and radial velocities for individual epochs.}             
\label{rvtab}      
\centering                          
\begin{tabular}{c c c  c c}        
\hline\hline                 
Date & MJD & Spectrum & Exposure time & $\rm RV_{h}$  \\    
\hline
   &   &  & (sec) &  (km/s)  \\
\hline                        
2016-02-19 & 57437.3686  &  {\rm GIRAFFE HR12}  & 1350 &      +227.4$\pm$0.2    \\
2016-02-19 & 57437.3414  &  {\rm GIRAFFE HR13}  & 1800 &      +227.4$\pm$0.2    \\
2016-02-19 & 57437.3155  &  {\rm GIRAFFE HR13}  & 1800 &      +227.7$\pm$0.2    \\
2016-02-01 & 57419.3284  &  {\rm GIRAFFE HR15N} & 2700 &      +228.0$\pm$0.1    \\
2016-02-12 & 57430.3564  &  {\rm GIRAFFE HR15N} & 2700 &      +227.8$\pm$0.1    \\
2016-02-01 & 57419.3671  &  {\rm GIRAFFE HR15N} & 2700 &      +228.2$\pm$0.1    \\
2017-06-17 & 57921.0096  &  {\rm X-SHOOTER}     & 1800 &      +235.7$\pm$0.3    \\      
2018-06-08 & 58277.1368  &  {\rm UVES}          & 1800 &      +226.90$\pm$0.05    \\
2018-06-08 & 58277.1586  &  {\rm UVES}          & 1200 &      +226.80$\pm$0.05    \\
2018-06-09 & 58278.1470  &  {\rm UVES}          & 1800 &      +227.00$\pm$0.05    \\
2018-06-27 & 58296.0161  &  {\rm HARPS}         & 3600 &      +226.90$\pm$0.05    \\    
2018-07-24 & 58323.9699  &  {\rm HARPS}         & 3600 &      +226.90$\pm$0.07    \\    
2019-02-21 & 58535.3547  &  {\rm HARPS}         & 3600 &      +227.90$\pm$0.05    \\    
2021-01-27 & 59241.3043  &  {\rm HARPS}         & 3600 &      +226.10$\pm$0.05    \\    

\hline                                   
\end{tabular}
\end{table*}

\section{Chemical abundances}

The UVES spectrum has been used to derive chemical abundances of Li, O, Na, Mg, Si, Ca, Ti, Fe and Ni, 
and to estimate the \ciso\ isotopic ratio, while the X-SHOOTER spectrum has been used 
to infer C, N, Al, Mg and Fe abundances. Solar reference abundances are from \citet{gs98} but for C, N and O 
taken from \citet{caffau11}. 

Abundances of Na, Mg, Si, Ca, Ti, Fe and Ni from the UVES spectrum have been obtained from the measured 
equivalent widths (EWs) using the code {\tt GALA}  \citep{gala}. 
EWs have been measured with the code {\tt DAOSPEC} \citep{stetson08} 
through the wrapper {\tt 4DAO} \citep{4dao}. 
All the other abundances have been derived by performing a $\chi^{2}$-minimisation with our proprietary code 
{\sl SALVADOR} between the observed spectrum and grids of 
synthetic spectra calculated with {\tt SYNTHE}.
One-dimensional, plane-parallel, LTE model atmospheres have been calculated with the code {\tt ATLAS9} \citep{kurucz05}.

Total uncertainties have been estimated by adding in quadrature the main sources errors, namely 
the errors arising from the measurement process and those arising from the adopted 
stellar parameters. 
For the elements derived from EWs, the internal error is computed as the line-to-line scatter 
divided by the root mean square of the number of lines. For the elements derived from spectral 
synthesis the uncertainty in the measure has been estimated with Monte Carlo simulations with the same approach described in Section 3.

Effective temperature (\teff ) and surface gravity (\gr ) of \#25664 have been estimated 
using the photometry of the early third data release (EDR3) of the ESA/Gaia mission \citep{prusti16,brown21}. 
We used a new implementation of the ${\rm (BP-RP)_0}$-\teff\ transformation by \citet{mb20gaia} 
based on the Gaia EDR3 photometry, adopting 
the colour excess E(B-V)=~0.12 mag \citep{harris10} and correcting for the extinction following 
the procedure by \citet{bab18}. The associated error, based on the uncertainty of the 
photometry, reddening and the adopted colour-\teff\ calibration is $\sim$90 K.
Surface gravity has been computed adopting the photometric \teff , 
a mass of 0.8 $\rm M_{\odot}$ and G-band bolometric corrections computed according to \citet{andrae18}. 
The microturbulent velocity has been estimated using the standard approach to minimise the trend between 
the abundance of Fe and the reduced EW.
We estimated \teff\ =~5116$\pm$90 K, \gr\ =~2.44$\pm$0.1 and $v_t$=~1.4$\pm$0.1 km/s. 

The analysis of the UVES spectrum  provides [Fe~I/H]=--1.74$\pm$0.09 dex and [Fe~II/H]=--1.69$\pm$0.05 dex, 
with an excellent match between the two measures of Fe abundances. 
The adopted \teff\ provides a negative slope (--0.03$\pm$0.01 dex/eV) between iron abundances and excitation potential 
that can be minimised decreasing \teff\ by about 200 K . This difference between spectroscopic 
and photometric \teff\ for metal-poor stars has been already discussed by \citet{mb20} who 
recommend to use the photometric \teff\ values even if they introduce negative slopes with 
the excitation potential.

The atmospheric parameters adopted by \citet{mu19} 
have been derived adopting optical and near-infrared ground-based photometry, and the colour-\teff\ transformations by \citet{alonso99}, 
obtaining \teff\ = 4958 K and \gr\ = 2.37. 
The \teff\ based on the Gaia EDR3 photometry and used in this work (\teff\ = 5116 K)
is $\sim$160 K hotter than that adopted by \citet{mu19}. 
This difference reflects mainly the intrinsic difference between the two 
colour-\teff\ transformations. 
The difference in the two sets of parameters leads to a variation in \fei\ , 
mainly sensitive to \teff\  , but  similar [Fe~II/H] , because 
the variation of [Fe~II/H] due to \teff\ compensates that due to \gr\ 
($\delta$\gr =+0.07). 
When the parameters by \citet{mu19} are adopted we derive 
\fei\ =--1.88$\pm$0.09 dex and \feii\ = --1.71$\pm$0.05 dex. 
The Gaia/EDR3 \teff\ provides a better agreement between the two Fe abundances.
On the other hand, the two sets of parameters lead to similar abundance ratios [X/Fe], 
with variations smaller than 0.05 dex.

Also for the two comparison stars observed with X-SHOOTER the atmospheric parameters 
have been re-derived using the Gaia EDR3 photometry, finding
\teff\ =~5076 K, \gr\ =~2.46 for the 1P star \#77093, and 
\teff\ =~5104 K, \gr\ =~2.49 for the 2P star \#329049.
For sake of homogeneity, [Fe/H] of the three stars has been estimated from about 10 unblended 
Fe~I lines from the X-SHOOTER spectra. For \#25664, [Fe/H] from UVES and X-SHOOTER well match 
within the uncertainties.

\subsection{Lithium}
The resonance Li line at 6707 \AA\ is detected in each 
spectrum because of its huge EW ($\sim$230 m\AA\ ). 
We derived the Li abundance from the UVES spectrum that 
allows to measure both the resonance line and the subordinate 
line at 6103 \AA\ (not visible in the other spectra due to its 
small strength). Lithium abundances from these two lines have been corrected 
for NLTE effects according to \citet{lind08}.

\begin{figure}[!ht]
\centering
\includegraphics[scale=0.40]{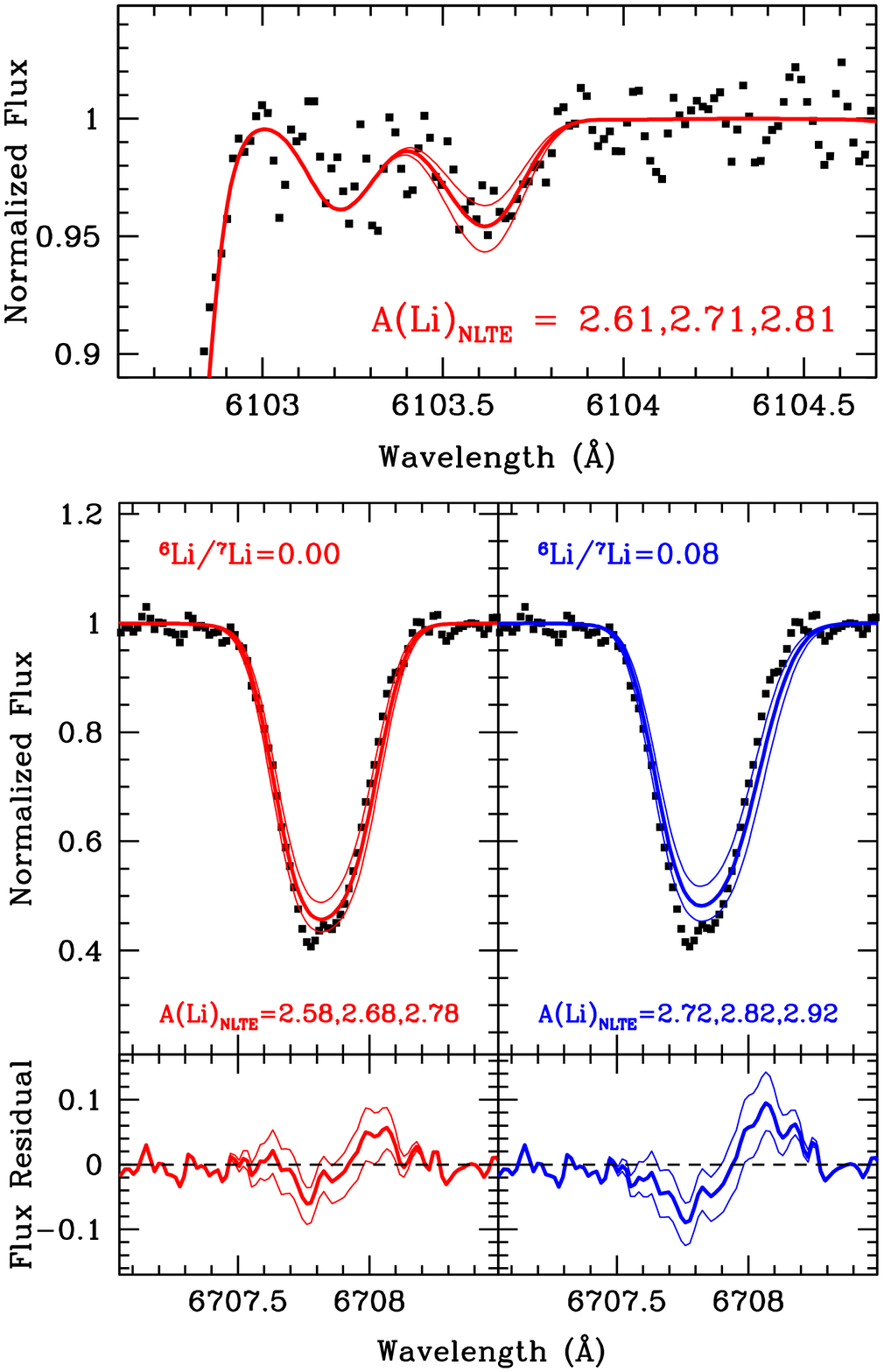}
\caption{Portions of the UVES spectrum (black squares) around the subordinate and 
resonance Li lines (upper and middle/lower panels, respectively). 
In the upper panel, the observed spectrum is superimposed to the best-fit synthetic 
spectrum (thick red line) and two synthetic spectra calculated with \ali\ $\pm$0.1 dex 
with respect to the best-fit one (thin red lines), assuming ${\rm ^{6}Li/^{7}Li}$=0.00 . 
In the middle panels, the observed spectrum is compared with a set of synthetic spectra 
calculated with the best-fit abundance (thick line) and \ali\ $\pm$0.1 dex 
with respect to the latter (thin lines), assuming ${\rm ^{6}Li/^{7}Li}$=0.00 and 0.08 
(left and right panels, respectively). The lower panels show the residuals between 
the observed spectrum and the synthetic spectra shown in the above panels.}
\label{fitlit}
\end{figure}

The best-fit of the resonance line in the UVES spectrum is not 
fully satisfactory, because at this spectral resolution 
we are not able to reproduce simultaneously the line depth and 
the line broadening, in particular the red wing of the line.  
These effects were not revealed with the GIRAFFE spectra because of their lower 
spectral resolution.
The lower panels of Fig.~\ref{fitlit} show the observed line profile compared to a grid of synthetic spectra computed with two different ${\rm ^{6}Li/^{7}Li}$ isotopic ratios, namely 0.0 and 0.08 (right and left panel, respectively). 
The assumption of no ${\rm ^{6}Li}$ (instead of a solar ${\rm ^{6}Li/^{7}Li}$  isotopic ratio) improves the quality of the fit,
but again we cannot reproduce the entire line profile. 
We ascribe this effect to the use of 1D model atmospheres 
that crudely describe the velocity fields in the stellar photospheres, with a relevant 
impact on strong and asymmetric lines like that at 6708 \AA\ .
The final Li abundance obtained by fitting the entire profile of the resonance Li line 
is  \ali\ =+2.68$\pm$0.10 dex assuming no $^{6}Li$.
That results to be an intermediate value between the abundance 
obtained fitting only the wings (\ali\ =+2.60 dex ) and excluding the 
red wing (\ali\ =+2.74 dex). 
The quoted uncertainty is dominated by the uncertainties in \teff\ , while the 
other parameters have a null impact on the abundance, and the error in the fitting procedure 
(estimated according to Monte Carlo simulations) is only 0.01 dex because of the 
high S/N ratio of the UVES spectrum and the intensity of the Li line.

The weak subordinate Li line at 6103 \AA\ provides a NLTE abundance 
of \ali\ =+2.71$\pm$0.08 dex (see upper panel of Fig.~\ref{fitlit}).
The error in the fitting procedure is $\pm$0.06 dex, 
higher than that estimated for the strong Li resonance line. On the other hand, 
this line has the advantages to be less sensitive to \teff\ with respect to the resonance line,
to be insensitive to the assumed ${\rm ^{6}Li/^{7}Li}$  isotopic ratio 
and to have a milder sensitivity to NLTE effects 
(the NLTE correction is +0.08 dex for the subordinate line and --0.22 dex for the 
resonance line).
For these reasons we decided to adopt the lithium abundance of the subordinate line.
 This value confirms the Li enhancement of this star, without changing our previous 
conclusions.

\subsection{Oxygen, Sodium, Magnesium and Aluminium}
Here we discuss the chemical abundances for the light elements 
(O, Na, Mg and Al) involved in the abundance patterns of globular cluster stars.\\
The only available indicator of O abundance is the forbidden 
oxygen line at 6300 \AA\ . Due to the relatively high \teff\ and the low metallicity of the star, 
the line is very weak also in the UVES spectrum despite its S/N ratio. 
Using synthetic spectra we can provide only an upper limit 
for the O abundance ([O/Fe]$<$+0.30 dex). 

\citet{mu19} derived the Na abundance of \#25664 
from the Na D lines in order to compare its [Na/Fe] with those 
of other LRGB stars in $\omega$ Centauri discussed in \citet{mu18}. 
The UVES spectrum analysed here allows us to measure the Na doublets at 
5682-88 \AA\ and at 6154-6160 \AA\ which are usually adopted as   
indicators of Na abundance in giant stars. From these lines we found a NLTE 
abundance [Na/Fe]=+1.00$\pm$0.05 dex, adopting the NLTE corrections 
by \citet{lind11}. This abundance is 0.13 dex higher than
the previous estimate based on Na D lines and the \citet{alonso99} \teff\ scale,  
confirming the extraordinary [Na/Fe] enhancement of this star.

\begin{figure} 
\centering
\includegraphics[clip=true,scale=0.45,angle=0]{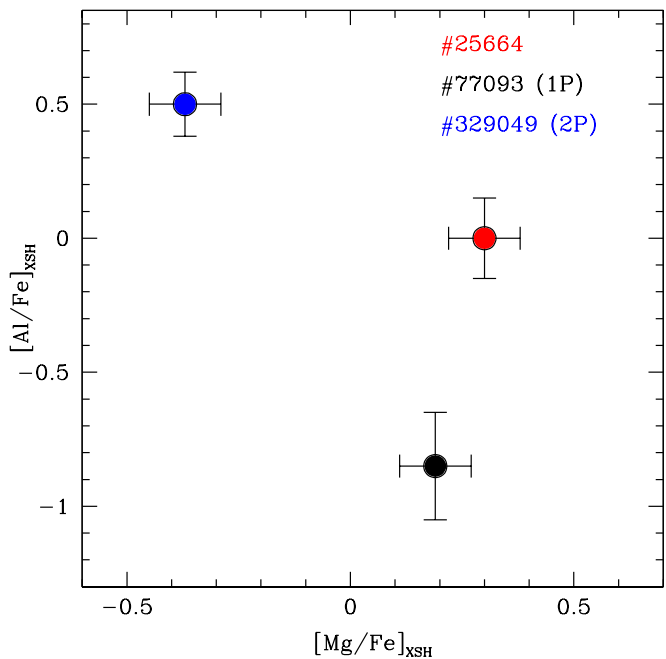}
\caption{Behaviour of [Al/Fe] as a function of [Mg/Fe] derived 
from the X-SHOOTER spectra for \#25664 and the two comparison stars.}
\label{mgal}
\end{figure}

From the X-SHOOTER spectrum we derived abundances of Al and Mg 
for the Li/Na-rich star and the two comparison stars.
The Al abundance cannot be inferred from the UVES spectrum because the 
Al doublet at 6696-98 \AA\ is too weak, while the X-SHOOTER 
spectra sample the resonance Al line at 3961 \AA\ . 
Additionally, we can measure for the three X-SHOOTER targets 
the Mg b triplet at 5165-5185 \AA\ . 
The Mg abundances in \#25664 derived using different lines 
in the UVES and X-SHOOTER spectra are in excellent agreement.
Figure~\ref{mgal} shows the behaviour of [Al/Fe] as a function of [Mg/Fe] 
for these three targets. 
The three stars are located in different parts of the [Mg/Fe]-[Al/Fe] diagram,
with \#25664 having [Mg/Fe] similar to the 1P star but 
[Al/Fe] intermediate between those measured in the two reference stars.

\subsection{C, N and \ciso }

C and N abundances have been derived from the comparison between synthetic spectra 
and the flux-calibrated X-SHOOTER and UVES spectra. 
C abundances have been obtained by fitting the CH G-band 
($A^{2}\Delta-X^{2}\Pi$) at $\sim$4300 \AA\ , employing the 
most recent linelist by \citet{masseron14}. 
N abundances have been derived using 
the NH ($A^{3}\Pi_{i}-X^{3}\Sigma^{-}$) molecular band at $\sim$3360 \AA ; 
additionally, we derived N abundances also from
the CN ($B^{2}\Sigma_{}-X^{2}\Sigma^{}$) molecular band 
at 3880 \AA\  (for both molecules we adopted the 
linelists available in the R. L. Kurucz database). 
The two N diagnostics provide different results 
\citep[see e.g.][]{spite05} but we privilege 
the NH band because it is a {\sl pure} indicator of the N abundances, 
while the CN band is also sensitive to the C abundance.  
For the CN band, we derived the N 
abundances by fixing the C abundance obtained from the CH band.

We prefer to use the flux-calibrated spectra in order to 
reduce the problems related to the spectral normalisation that can be 
a critical issue in heavily blanketed regions. 
For each wavelength the flux has been corrected
for the reddening adopting the extinction law by \citet{sm79} 
and a colour excess E(B-V)=~0.12 mag \citep{harris10}. 
The dereddened flux-calibrated spectra have then been compared with synthetic spectra. 
In this way, we need only to apply a scaling factor 
to match observed and synthetic spectra, avoiding the risk to distort 
the real shape of the observed spectrum 
with a potential significant impact on the derived C and N abundances.
The C and N abundances of \#25664 derived from X-SHOOTER and UVES spectra 
are in excellent agreement. In the following we refer to the C and N abundances 
from X-SHOOTER to compare the abundances of \#25664 with those of the two 
reference stars observed with X-SHOOTER.

\#25664 is enhanced both in C and in N abundances
(the latter regardless of the diagnostic, with the CN band that provides 
N abundances higher than the NH band by $\sim$0.4-0.5 dex). 
The two comparison stars show different patterns, as expected: 
the C abundances in the two stars are similar, 
with a slightly lower C abundance in the 2P star. The latter exhibits a strong enhancement in N 
with respect to 1P star. 
When we compare \#25664 with these two stars
we find that it has [N/Fe] compatible with that of 
the 2P star, while its [C/Fe] is higher than those of both stars.

\begin{figure*} 
\centering
\includegraphics[width=17cm]{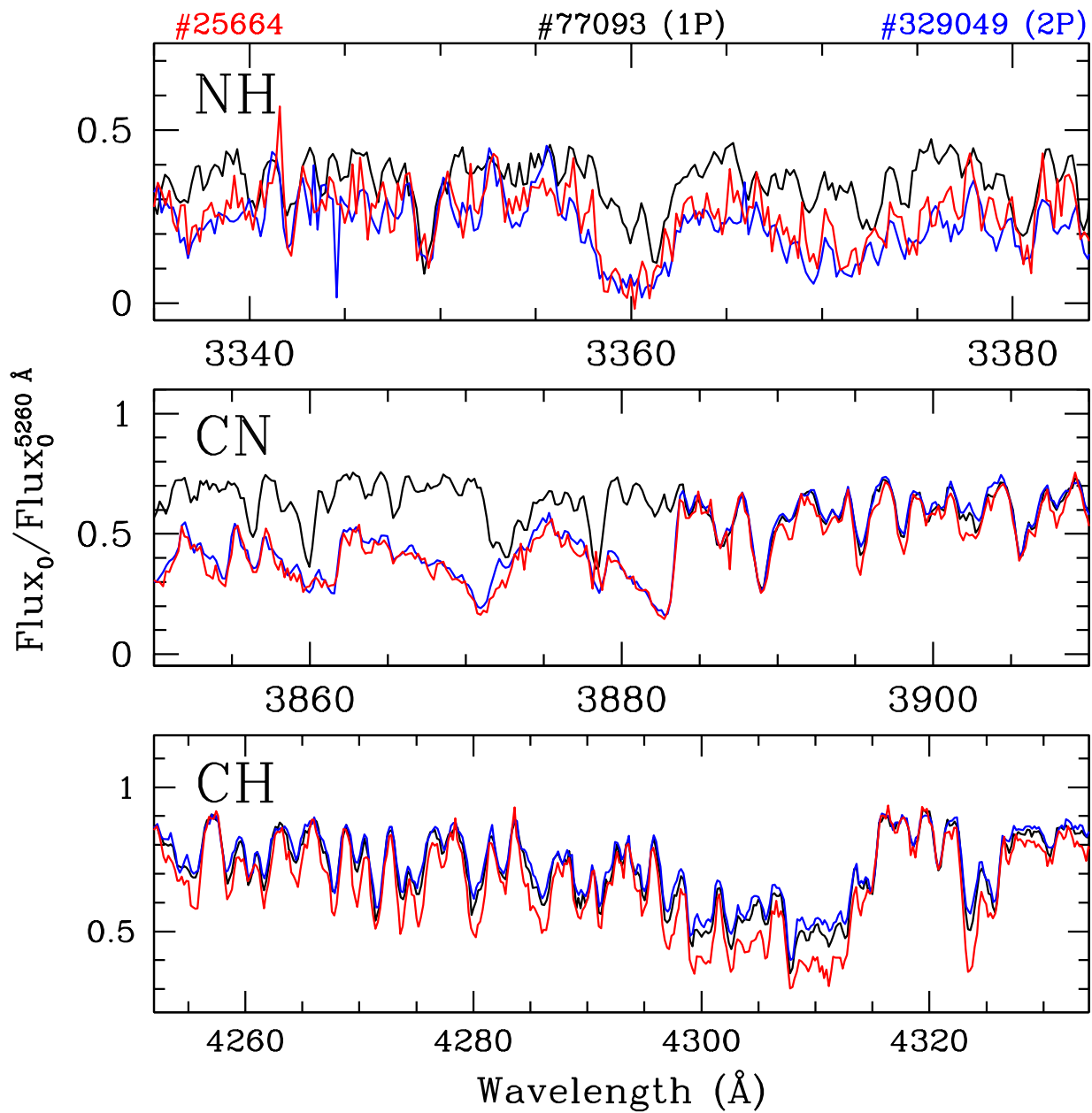}
\caption{Comparison among the X-SHOOTER spectra of \#25664 (red line) and the 
two reference stars (black line for 1G and blue line for 2G) for the 
NH (upper panel), CN (middle panel) and CH (lower panel) molecular bands. 
The flux-reduced spectra are corrected for extinction and normalised to the flux measured at 5260 \AA\ .}
\label{band1}
\end{figure*}

The comparison between NH, CN and CH bands in these three stars is displayed in Fig.~\ref{band1}, 
showing both the similar N abundance between \#25664 and the 2P reference star,  
and the higher C abundance of \#25664 with respect to the two reference 
stars. 

Adopting the C abundance obtained with the X-SHOOTER spectrum, we measured 
the \ciso\ isotopic ratio from about 10 $^{13}{\rm CH}$ features in the UVES 
Blue Arm flux-calibrated spectrum, finding an average value of 15$\pm$2. 
Giant stars affected by further mixing after the first dredge-up
have values of \ciso\ around 5-6, while metal-poor LRGB stars 
have values higher than 13-15 \citep[see e.g.][]{spite06}.
The value measured in \#25664 points out that this star 
has not yet undergone the extra-mixing episode usually 
associated with the RGB bump.

\begin{table}
\caption{Chemical abundances of \#25664 from the UVES spectrum.}             
\label{tabuves}      
\centering                          
\begin{tabular}{c c  }        
\hline\hline                 
Ion &   Abundance  \\    
\hline                        

 ${\rm [Fe~I/H]}$   &     --1.74$\pm$0.09    \\
 ${\rm [Fe~II/H]}$  &     --1.69$\pm$0.05    \\
  \ali  &      +2.71$\pm$0.07    \\
 ${\rm [O/Fe]}$  &      $<$+0.30    \\
${\rm [Na/Fe]_{NLTE}}$  & +1.01$\pm$0.05    \\
${\rm [Mg/Fe]}$  &        +0.30$\pm$0.04    \\
${\rm [Si/Fe]}$  &        +0.38$\pm$0.05    \\
${\rm [Ca/Fe]}$  &        +0.30$\pm$0.03    \\
${\rm [Ti~I/Fe]}$  &       +0.20$\pm$0.03   \\
${\rm [Ti~II/Fe]}$  &      +0.33$\pm$0.08    \\
${\rm [Ni/Fe]}$  &      --0.04$\pm$0.03    \\

\hline                                   
\end{tabular}
\end{table}

\begin{table}
\caption{Chemical abundances of \#25664 and of the comparison stars 
\#77093 and \#320949 from the X-SHOOTER spectra.}             
\label{tabxsh}      
\centering                          
\begin{tabular}{c c c c }        
\hline\hline                 
 & \#25664 & \#77093 & \#329049  \\    
 &       & (1P) & (2P)  \\    
\hline                        
 ${\rm [Fe/H]}$      &  --1.67$\pm$0.13  &  --1.62$\pm$0.12   & --1.65$\pm$0.14 \\
 ${\rm [C/Fe]}$      &  +0.45$\pm$0.16  &  --0.08$\pm$0.16   & --0.15$\pm$0.15 \\
 ${\rm [N/Fe]_{NH}}$ &  +0.99$\pm$0.20  & --0.68$\pm$0.20   & +1.10$\pm$0.20 \\
 ${\rm [N/Fe]_{CN}}$ &  +1.23$\pm$0.14  & $<$0.0        & +1.60$\pm$0.15   \\
 ${\rm [Mg/Fe]}$     &  +0.30$\pm$0.08  &  +0.19$\pm$0.08   & --0.37$\pm$0.08 \\
 ${\rm [Al/Fe]}$     & +0.00$\pm$0.15  & --0.85$\pm$0.20   &  +0.50$\pm$0.12\\
 ${\rm [K/Fe]}$      &  +0.30$\pm$0.10  & +0.19$\pm$0.11    & +0.25$\pm$0.08 \\

\hline                                   
\end{tabular}
\end{table}

\subsection{Potassium}       

Two GCs, namely NGC~2419 and NGC~2808, exhibit intrinsic scatter in 
their potassium abundances \citep{mu12,cohen12,mu15}. The enhancement of [K/Fe] is coupled with 
significant depletion of [Mg/Fe], defining a clear Mg-K anticorrelations. 
In other GCs K abundances have only small or null [K/Fe] spreads \citep{carretta13,mu17,cerni17,cerni18}. 

We derived potassium abundances in the three X-SHOOTER spectra from the resonance line at 7699 \AA\ . 
This transition suffers from significant NLTE corrections. The three stars have similar stellar parameters, 
metallicity and LTE K abundance, therefore they should have  the same NLTE correction for the K resonance line.
According to the NLTE calculations presented by \citet{mu17} we applied to the K abundances of the three stars 
an offset of --0.35 dex. No significant difference in the K abundances is found.

\section{Discussion}

The multi-instrument spectroscopic analysis 
of the peculiar LRGB star \#25664 of $\omega$ Centauri provides the following results:
\begin{itemize}

\item we confirm that the star is characterised by an extraordinary high abundance both of Li and Na, 
providing abundances based on diagnostics less sensitive to 
the ${\rm ^{6}Li/^{7}Li}$ isotopic ratio, 
NLTE and/or 3D effects with respect to those used by\citet{mu19}. Once the effect of the 
first dredge up is accounted for, the initial A(Li) of this star is predicted to be around 4.0~dex.

\item the RVs measured in different epochs point out a RV variability of this star, 
suggesting that it could be member of a binary system;

\item the star has not yet experienced the extra-mixing process occurring at the 
RGB bump luminosity level, as demonstrated by the measured \ciso\ isotopic ratio;

\item \#25664 is enriched both in [C/Fe] and [N/Fe]. [N/Fe] is compatible 
with the value measured in the 2P comparison star, while [C/Fe] is higher 
than both 1P and 2P comparison stars;

\item the star has enhanced [Mg/Fe] and [Al/Fe] abundance ratios. 
In particular the latter turns out to be intermediate between the Al abundances of the 1P and 2P 
reference stars. 

\item the other chemical species measured here do not highlight oddities, pointing out that 
the star is enhanced in [$\alpha$/Fe] abundance ratios (Si, Ca, Ti) like the other 
stars of $\omega$ Centauri of similar metallicity \citep[see e.g.][]{johnson10}.

\end{itemize}

This analysis confirms the uniqueness of the star \#25664, which is
distinct from the other Li-rich GC stars known so far, not only for its over-abundances 
of Li and Na but also for its peculiar combination of high C and N abundances. 
In terms of C and N, \#25664 does not resemble either the 1P or the 2P reference stars that we measured.
In fact, among the GC stars we should expect a more or less pronounced C-N anticorrelation or at least a large spread in N abundance at almost constant C abundance \citep[see e.g.][]{lardo12,lardo13,yong15,holly17}, 
the latter case similar to what we observe for the two reference stars. 
Even if the complex chemical patterns of $\omega$ Centauri in terms of C and N abundances cannot be captured by these 
two reference stars only, this comparison reveals 
that \#25664 has been enriched in C and N in an anomalous way compared 
to what we expect, i.e. similar or lower C abundances in stars with enhanced N abundances.

In one of the possible scenarios discussed by \citet{mu19}, 
the circulation due to the onset of extra-mixing
processes (in addition to convection) after the RGB bump should be able to 
move $^{3}{\rm He}$ from the convective envelope down to 
regions hot enough to form $^{7}{\rm Be}$ which must be transported back to the envelope 
to form $^{7}{\rm Li}$ \citep[see][]{dv03}.  
The  \ciso\ isotopic ratio of \#25664 (15$\pm$2) is consistent with the values measured 
in pre-RGB bump stars.
This shows that \#25664 has not yet experienced the extra-mixing episode occurring at 
the RGB bump, because mixed stars have \ciso\ around 5-6 and in general well below 
10 \citep{gratton00,spite06}. 
Therefore this scenario of internal Li production can be ruled out.

Other two viable scenarios can be envisaged to explain the anomalous chemical pattern 
of \#25664, namely a mass transfer process within a binary system or the formation of this star 
from the pure ejecta of super-AGB stars.
In the first case \#25664  could be member of a binary system with a companion star, 
that is now a faint compact object, most likely a white dwarf. 
This scenario is favoured by the RV variability detected with our multi-epoch spectroscopic dataset. 
During its main sequence evolution \#25664 could have experienced
mass accretion from the companion's progenitor, and the accreted gas must have been 
then diluted in its convective envelope to a degree that depends on the exact value 
of the initial mass of \#25664, the timing and amount of accreted gas.

According to the theoretical models for the AGB ejecta \citep{ventura13,doherty14} 
we identify two potential candidates for the companion star's progenitor, 
namely AGB stars with 3-4 and 7-8 $M_{\odot}$, because their winds 
are predicted to contain high abundances of Li and Na. 
However, in order to properly compare the measured abundances with the predicted values, 
appropriate models for the evolution of the binary system should be computed accounting for the efficiency, the timing and the duration of the 
accretion. 

At our request Pasquale Panuzzo (GEPI, France) did a bayesian analysis of the
RV measurements. The analysis displays peaks in the posterior probability distribution of periods at around 550, 770 and 1130 days, the latter being the maximum peak. A unique determination of the period would requires further
observations, with a better coverage of orbital phases.

Another fascinating scenario to explain \#25664 has been proposed by \citet{mu19}, 
suggesting that this star formed directly from the pure ejecta of a super-AGB star before the dilution 
with the pristine gas \citep{dantona12}.
The existence in some clusters of a small fraction of stars formed through this process 
has been mainly proposed to explain the He-rich (Y$\sim$0.35) stellar population in some 
clusters, including Omega Centauri.

The new chemical and kinematical data presented in
this work make more intriguing to explain the peculiar
surface chemistry of \#25664, that does not resemble any other
GC star observed so far and remains {\sl a unicum} among 
the GC stars. 
New RV measures, performed with a suitable time sampling, are needed to 
firmly confirm or reject the binary scenario. 
For both the scenarios new, dedicated theoretical models are needed 
in order to properly reproduce the entire chemical composition of \#25664.

\begin{acknowledgements}
We thank the referee, Franca D'Antona, for the useful comments and suggestions. 
We are extremely grateful to Pasquale Panuzzo for analysing our radial velocity data with his bayesian code {\sl Batman}.
This work has made use of data from the European Space Agency (ESA) mission
{\it Gaia} (\url{https://www.cosmos.esa.int/gaia}), processed by the {\it Gaia}
Data Processing and Analysis Consortium (DPAC,
\url{https://www.cosmos.esa.int/web/gaia/dpac/consortium}). Funding for the DPAC
has been provided by national institutions, in particular the institutions
participating in the {\it Gaia} Multilateral Agreement.
\end{acknowledgements}

\bibliographystyle{apj}
{}

\end{document}